\documentclass[a4paper,11pt]{article}
\pdfoutput=1
\usepackage{jheppub}

\usepackage{preprintcover} 
\PreprintCoverPaperTitle{Measurement of the differential cross-section of
  $B^+$ meson production in $\pp$ collisions at $\sqrt{s}=7$~TeV
  at ATLAS} 
\PreprintIdNumber{CERN-PH-EP-2013-089}
\PreprintCoverAbstract{The production cross-section of $B^+$ mesons is measured as a
  function of transverse momentum $\pt$ and rapidity $y$ in
  proton--proton collisions at centre-of-mass energy $\sqrt{s}=7$\;TeV, using 2.4 fb$^{-1}$ of data recorded
  with the ATLAS detector at the Large Hadron Collider.
  The differential production cross-sections, determined in the range
  $9$\;GeV\;$<\pt<120$\;GeV and $|y|<2.25$, are compared to
  next-to-leading-order theoretical predictions.}
\PreprintJournalName{JHEP}

\usepackage[T1]{fontenc} 

\usepackage{lscape}
\usepackage{atlasphysics}
\usepackage{multirow} 
\usepackage{subfigure}
\RequirePackage{xspace}
\usepackage{amsmath}
\usepackage{calrsfs} 

\DeclareMathAlphabet{\pazocal}{OMS}{zplm}{m}{n}

\usepackage{rotating}
\usepackage{perpage}
\usepackage{upgreek}
\usepackage{hyperref}
\MakePerPage{footnote}

\title{\boldmath Measurement of the differential cross-section of
  $B^+$ meson production in $\pp$ collisions at $\sqrt{s}=7$~TeV
  at ATLAS}

\author{The ATLAS Collaboration}

\abstract{The production cross-section of $B^+$ mesons is measured as a
  function of transverse momentum $\pt$ and rapidity $y$ in
  proton--proton collisions at centre-of-mass energy $\sqrt{s}=7$\;TeV, using 2.4 fb$^{-1}$ of data recorded
  with the ATLAS detector at the Large Hadron Collider.
  The differential production cross-sections, determined in the range
  $9$\;GeV\;$<\pt<120$\;GeV and $|y|<2.25$, are compared to
  next-to-leading-order theoretical predictions.}

\begin{document} 
\maketitle
\flushbottom

\section{Introduction}
\label{section:intro}

The $b$-hadron production cross-section has been predicted with next-to-leading-order (NLO) accuracy for more than twenty years~\cite{nlo1,nlo2} and more recently it has been predicted with fixed order plus next-to-leading-logarithms (FONLL) calculations~\cite{fonll,fonll2}. Several measurements were performed with proton--antiproton collisions by the UA1 collaboration at the S$p\bar{p}$S collider (CERN) at a centre-of-mass energy of $\sqrt{s} = 630$\;GeV~\cite{ua1,ua2} and by the CDF and D0 collaborations at the Tevatron collider (Fermilab) at $\sqrt{s} = 630$\;GeV, 1.8\;TeV and 1.96\;TeV~\cite{cdf0,cdf1,cdf2,cdf3,d01,d02,d03,cdf4,cdf5,cdf6}.
These measurements made a significant contribution to the understanding of heavy-quark production in hadronic collisions~\cite{cacciari}.
However, the dependence of the theoretical predictions for $b$-quark production on the factorisation and renormalisation scales and the $b$-quark mass $m_b$~\cite{nlo2} results in theoretical uncertainties of up to 40\% and, therefore, it is important to perform precise measurements of $b$-hadron production cross-sections.
In addition, measurements of $b$-hadron production cross-sections are of theoretical interest at higher $\sqrt{s}$~\cite{nlo3} and for $B$ mesons of higher transverse momentum ($\pt$)~\cite{fonll2012}.

Measurements of the $b$-hadron production cross-section in proton--proton collisions at the
Large Hadron Collider (LHC) provide further tests of QCD calculations for heavy-quark
production at higher centre-of-mass energies and in wider transverse momentum ($\pt$) and rapidity ($y$) ranges, thanks to the extended coverage and excellent performance of the LHC detectors.
Recently the LHCb collaboration measured $b$-hadron production cross-sections using $B^\pm\to\Jpsi\,K^\pm$, $b\to\Jpsi\,X$ and semileptonic $b$-hadron decays in the forward rapidity region at $\sqrt{s} = 7$\;TeV~\cite{lhcb1,lhcb2,lhcb3}. The CMS collaboration measured the production cross-sections for $B^+$, $B^0$, $B_s$ mesons, $\Lambda_b$ baryons, and inclusive $b$-hadron production 
using $b\to\Jpsi\,X$ decays, semileptonic decays, and $b$-hadron jets at $\sqrt{s} = 7$\;TeV~\cite{cms1,cms2,cms3,cms4,cms5,cms6,cms7,cms8}. ATLAS has measured $b$-hadron production cross-sections using semileptonic decays~\cite{DstarMuX,inclLept}, $b\to\Jpsi\,X$ decays~\cite{Jpsi} and $b$-hadron jets~\cite{bjets}. 

This paper presents a measurement of the $B^+$ production cross-section using the decay channel $B^+\to J/\psi K^+\to\mu^+\mu^-K^+$ in $pp$ collisions at $\sqrt{s} = 7$\;TeV, as a function of $B^+$ transverse momentum and rapidity. 
The ATLAS and CMS detectors provide coverage in the central rapidity region, so their measurements are complementary to the LHCb measurements. With 2.4\;fb$^{-1}$ of data collected by the ATLAS detector, this analysis extends the measurement of the $B^+$ cross-section up to $\pt$ of about 100\;GeV, allowing comparisons with NLO predictions in four rapidity regions in the range $|y|<2.25$ to be made.
The results are reported for $B^+$ meson production, but are derived from both charged states, under the assumption that in the phase space accessible by this measurement the $B^+$ and $B^-$ production cross-sections are equal.
This assumption is in agreement with the predictions of NLO Monte Carlo generators and is also valid
within the precision of the measurement.

\section{The ATLAS detector}

The ATLAS experiment~\cite{ATLAS} uses a general-purpose detector
\footnote{ATLAS uses a right-handed coordinate system ($x,y,z$) with its origin at the nominal interaction point. The $z$-axis is along the beam pipe, the $x$-axis points to the centre of the LHC ring and the $y$-axis points upward. Cylindrical coordinates ($r$, $\phi$) are used in the transverse plane, $\phi$ being the azimuthal angle around the beam pipe. The pseudorapidity $\eta$ is defined as $\eta = -\ln[\tan(\theta/2)]$, where $\theta$ is the polar angle.} consisting of an inner tracker, a calorimeter and a muon spectrometer.
A brief outline of the components that are most relevant for this analysis is given below.
The inner detector (ID) directly surrounds the interaction point; it includes a silicon pixel detector (Pixel), a silicon microstrip detector (SCT) and a transition radiation tracker (TRT), and is embedded in an axial 2\;T magnetic field.  The ID covers the range $|\eta|<2.5$ and is enclosed by a calorimeter system containing electromagnetic and
hadronic sections.
The calorimeter is surrounded by a large muon spectrometer (MS) inside an air-core toroidal magnet system that contains a combination of monitored drift tubes
(MDTs) and cathode strip chambers (CSCs), designed to provide precise
position measurements in the bending plane in the range $|\eta|<2.7$.
In addition, resistive plate chambers (RPCs) and thin gap chambers (TGCs) with a coarse position resolution but 
a fast response time are used primarily to trigger muons in the ranges $|\eta|<1.05$ and $1.05 <|\eta|< 2.4$, respectively. RPCs and TGCs are also used to provide position measurements in the non-bending plane and to improve pattern recognition and track reconstruction. Momentum measurements in the MS are based on track segments formed in at least two of the three stations of the MDTs and the CSCs.

The ATLAS trigger system~\cite{ATLAStrigger} has three levels: the hardware-based Level-1 trigger and the two-stage High Level Trigger (HLT), comprising the Level-2 trigger and Event Filter (EF). At Level-1, the muon trigger searches for patterns of hits satisfying different $\pt$ thresholds using the RPCs and TGCs. The region-of-interest (RoI) around these Level-1 hit patterns then serves as a seed for the HLT muon reconstruction, in which dedicated algorithms are used to incorporate information from both the MS and the ID, achieving a position and momentum resolution close to that provided by the offline muon reconstruction.

\section{Data and Monte Carlo samples}
\label{sec:dataAndMC}

The analysis is based on data collected at the LHC during the proton--proton running period in the early 2011 (April--August) with a dimuon trigger that required the presence of at least two muon candidates with $\pt>4$\;GeV. Later run periods are not considered because this trigger was prescaled.
Selected events are required to have occurred during stable LHC beam conditions and the ID, as well as the MS, must have been fully operational. The collected data correspond to an integrated luminosity of $2.4$~fb$^{-1}$ with an uncertainty of 1.8\%~\cite{Lumi}.

In the analysis two Monte Carlo (MC) samples are used. The first sample simulates the signal $B^{\pm}\to \Jpsi\,K^{\pm}\to\mu^+\mu^-\,K^{\pm}$, while the second simulates $b\bar{b}$ production, with $b\bar{b}\to\Jpsi\,X\to\mumu\,X$, including the signal and also the backgrounds which are relevant for the analysis.
Both samples were generated with \Pythia~6~\cite{Pythia6} using the 2011 ATLAS tune~\cite{MC11}. The response of the
ATLAS detector was simulated~\cite{AtlasSim} using \Geant4~\cite{Geant}.
Additional $\pp$ interactions in the same and nearby bunch crossings (pile-up) were included in the simulation.

The MC samples are used in several parts of the analysis. The first is the extraction of the fit models for signal and background. The second is the construction of efficiency maps for the muon trigger and reconstruction.
The third is the estimation of the signal reconstruction efficiency and the kinematic acceptance of the selection criteria applied to the final-state particles in each $\pt$ and rapidity interval used in the analysis.
In the MC samples generated with \Pythia, the decay $\Jpsi\to\mumu$ is isotropic. In order to take into account that the $\Jpsi$ meson is produced with zero helicity in the $B^\pm$ rest frame, in the analysis a weight proportional to $\sin^2{\theta^*}$ is applied to each event, where $\theta^*$ is the $\mu^+$ angle relative to the $B^\pm$ direction in the $\Jpsi$ rest frame.

To compare the cross-section measurements with theoretical predictions, NLO QCD calculations matched with a leading-logarithmic parton shower MC simulation are used. Predictions for $b\bar{b}$ production are evaluated with two packages: \POWHEG-\textsc{hvq}~(\POWHEG-\textsc{Box}~1.0)~\cite{pow1,pow2} and \MCatNLO~4.01~\cite{mcnlo1,mcnlo2}. For the hadronisation process, \POWHEG~is matched with \Pythia, which uses the
Lund string model~\cite{pythiafragm1} with the Bowler modification~\cite{pythiafragm2} of the Lund symmetric fragmentation function~\cite{pythiafragm3}. \MCatNLO~is matched with \HERWIG~\cite{herwig1}, which uses a cluster model for hadronisation~\cite{herwigfragm}. The $b$-quark production cross-section is also calculated in the FONLL theoretical framework~\cite{fonll2012}, permitting direct comparison with the data assuming the world average of the hadronisation fraction $f_{\bar{b}\to B^+} = 0.401\pm 0.008$~\cite{PDG2012}. The theoretical uncertainties associated with the \POWHEG, \MCatNLO~and FONLL predictions are discussed in section \ref{sec:results} where the comparisons to the measured cross-sections are made.

\section{Event selection and reconstruction}
\label{sec:evsel}

Events for the analysis were selected with a trigger that requires two
muon RoIs at Level-1.
A full track reconstruction of dimuon candidates was performed by
the HLT where both muons are required to have $\pt>4$\;GeV and
fullfill additional requirements, loosely selecting events compatible
with $\Jpsi$ meson decays into a muon pair.

Events selected by the trigger are required to have at least one
reconstructed primary vertex with a minimum of three associated tracks.
Tracks reconstructed in the ID which are matched to tracks
reconstructed in the MS are selected as muon candidates.
Muon candidates are required to have sufficient numbers of
hits in the Pixel, SCT and TRT detectors to ensure accurate ID
measurements. The same selection criteria are applied to tracks
selected as potential $K^\pm$ candidates.

Events are required to contain at least one pair of reconstructed
oppositely signed muons that fit successfully to a common vertex,
using a vertexing algorithm~\cite{VKalVrt}.
The momenta of the muons and the dimuon invariant mass are calculated
from the refitted track parameters returned by the vertexing
algorithm.
Muon pairs with a common vertex are considered as
$J/\psi\to\mu^{+}\mu^{-}$ candidates if their invariant mass lies in
the mass range 2.7--3.5\;GeV.
Because of the trigger requirements on muons, the reconstructed
$\Jpsi$ candidate must have rapidity $|y|<2.25$ and the reconstructed muons
$\pt>4$\;GeV and $|\eta|<2.3$.
To ensure that the muon pair from the $\Jpsi$ candidate is the
one that triggered the event, an $(\eta,\phi)$ match between the trigger muons and those
of the $\Jpsi$ candidate is required. 
If multiple $\Jpsi$ candidates are found in the event, all are considered in the formation of $B^\pm$ candidates.

The muon tracks of the selected $J/\psi$ candidates are again fitted to a
common vertex with an additional third track with $\pt$ greater than
1 GeV. 
The three-track vertex fit is performed by constraining the muon
tracks to the $J/\psi$ mass~\cite{PDG2012}. The $K^\pm$
mass is assigned to the third track and the $\mumu K^\pm$ invariant
mass is calculated from the refitted track parameters returned by the
vertexing algorithm.
Regarding the quality of the three-track vertex fit, the $\chi^2$ per degree of freedom
must be $\chi^2/N_{\textrm{d.o.f.}}<6$, which is found to select about
99\% of signal events while rejecting background events.
We retain $B^+$ and $B^-$ candidates with
$\pt>9$\;GeV and $|y|<2.25$ in the mass range
5.040--5.800\;GeV. After this selection, the 
average candidate multiplicity is 1.3. 
The multiple $B^\pm$ candidates result mainly from random combinations
of tracks with selected $\Jpsi$ mesons produced promptly in $\pp$
collisions.
Such combinations result in non-resonant background and do not affect the estimation of the
signal yield. 

\section{Cross-section determination} 
\label{section:crosssection}

The differential cross-section for $B^+$ meson production in $pp$ collisions times branching ratio to the final state is given by
\begin{equation}
  \frac{\textrm{d}^2\sigma (pp\to\Bplus X)}{\textrm{d}\pt \textrm{d}y}
  \cdot {\mathcal{B}}=
  \frac{N^{B^+}}{{\cal{L}}\cdot\Delta \pt \cdot \Delta
    y} ~,
  \label{eq:xsec}
\end{equation}
where $\cal{B}$ is the total branching ratio of the signal decay,
which is $(6.03\pm 0.21)\times 10^{-5}$, obtained by combining the world-average values of the branching
ratios for $B^+\to\Jpsi\,K^+$ and $\Jpsi\to\mu^+\mu^-$~\cite{PDG2012},
$N^{B^+}$ is the number of $B^+\to\Jpsi\,K^+$ signal decays produced,
$\cal{L}$ is the integrated luminosity of the data sample and $\Delta
\pt$, $\Delta y$ are the widths of $\pt$ and $y$ intervals.
Assuming that $B^+$ and $B^-$ mesons are produced in equal numbers,
$N^{B^+}$ is derived from the average yield of the two
reconstructed charged states in a $(\pt, y)$ interval, after correcting for
detector effects and acceptance,
\begin{equation}
  N^{B^+} =
  \frac{1}{A}\frac{N^{B^+}_{\textrm{reco}}}{\varepsilon^{B^+}} = \frac{1}{A}\frac{N^{B^-}_{\textrm{reco}}}{\varepsilon^{B^-}} =
  \frac{1}{A}\frac{N^{B^\pm}_{\textrm{reco}}}{\varepsilon^{B^+}+\varepsilon^{B^-}}
  ~,
  \label{eq:NBplus}
\end{equation}
where $N^{B^\pm}_{\textrm{reco}}$ is the number of reconstructed
signal events, obtained from data with a fit to the invariant mass
distribution of $B^\pm$ candidates, $A$ is the acceptance of the
kinematic selection of the final-state particles of the signal decay,
obtained from MC simulation, and $\varepsilon^{B^{+}}$, $\varepsilon^{B^{-}}$ are the
reconstruction efficiencies for the $B^{\pm}$ signal
decays.
Separate efficiency is needed for $B^{+}$ and $B^{-}$ signal decays, because
the different interaction cross-sections of $K^{+}$ and $K^{-}$ with
the detector material result in different reconstruction efficiencies
for the two charged mesons.
The reconstruction efficiencies for $B^{+}$ and $B^{-}$ are obtained from MC
simulation.
In the following, $\varepsilon^{B^-}$ is implicitly referred to,
together with $\varepsilon^{B^+}$.
The efficiency for $B^{+}$ events is defined as the product of
trigger, muon reconstruction (ID and MS), kaon reconstruction and
vertexing efficiencies,
\begin{equation*}
  \varepsilon^{B^+} = \varepsilon_{\textrm{trigger}}^{\Jpsi}\cdot
  \varepsilon^{\mu^+} \cdot \varepsilon^{\mu^-} \cdot \varepsilon_{\textrm{ID}}^{K^+}
  \cdot \varepsilon_{\textrm{vertex}}^{\mu\mu K} =
  \varepsilon_{\textrm{trigger}}^{\Jpsi}\cdot
  \varepsilon_{\textrm{MS}}^{\mu^+} \cdot \varepsilon_{\textrm{MS}}^{\mu^-} \cdot
  (\varepsilon_{\textrm{ID}}^{\mu})^2 \cdot \varepsilon_{\textrm{ID}}^{K^+}
  \cdot \varepsilon_{\textrm{vertex}}^{\mu\mu K} ~.
  \label{eq:signalefficiency}
\end{equation*}
In the above equation, $\varepsilon_{\textrm{MS}}^{\mu^+}$ and $\varepsilon_{\textrm{MS}}^{\mu^-}$ are the
efficiencies for reconstructing $\mu^+$ and $\mu^-$ in the MS, which
differ for muons of low $\pt$ and large $|\eta|$ because of the bending
of tracks in the toroidal magnetic field.
This effect is to large extent symmetric for a simultaneous change of sign in the muon charge and in $\eta$.
The trigger efficiency, $\varepsilon_{\textrm{trigger}}^{\Jpsi}$,
depends on the ability of the trigger to identify muons of given $\pt$
and $\eta$ as decay products of a $\Jpsi$ meson.
The trigger efficiency includes independent and correlated terms
between the two muons\;\cite{Upsilon}.
The efficiency $\varepsilon^{B^+}$ for a given $(\pt, y)$ interval is
obtained from MC-simulated signal events from the fraction
\begin{equation}
  \varepsilon^{B^+} = \frac{N^{B^+}_{\textrm{\textsc{mc},reco}}}{N^{B^+}_{\textrm{\textsc{mc}, gen}}} ~, 
  \label{eq:eff}
\end{equation}
where the denominator is the number of signal events generated in a given
interval of the generated $\pt$ and $y$ and the numerator is the number of signal
events that pass the trigger and the offline selection requirements in
the same $(\pt, y)$ interval of the reconstructed variables.
Bin-to-bin migration effects are included in the efficiency definition
of eq.\;\eqref{eq:eff}.
The trigger and muon reconstruction efficiencies are measured in the
data using auxiliary single muon and dimuon triggers and tag-and-probe
methods\;\cite{Upsilon} and the simulation is corrected with per-event
weights to reproduce the efficiencies measured with data.
The derived weights, $w_{\textrm{MS}}^{\mu^+}$,
$w_{\textrm{MS}}^{\mu^-}$, $w_{\textrm{trigger}}^{\Jpsi}$, are applied
in each reconstructed MC-simulated signal event, so that $N^{B^+}_{\textrm{\textsc{mc}, reco}}$ in eq.\;\eqref{eq:eff} is now defined as
\begin{equation}
  N^{B^+}_{\textrm{\textsc{mc}, reco}} = \sum\limits_{i=1}^{N_{\textrm{\textsc{mc}}}^{\textrm{events}}} (w_{\textrm{MS}}^{\mu^+})_i \cdot (w_{\textrm{MS}}^{\mu^-})_i
  \cdot (w_{\textrm{trigger}}^{\Jpsi})_i ~,
  \label{eq:weights}
\end{equation}
where $N_{\textrm{\textsc{mc}}}^{\textrm{events}}$ is the number of
reconstructed MC-simulated signal events before applying the weights
derived from data.
The efficiency for reconstructing muons in the ID,
$\varepsilon_{\textrm{ID}}^\mu$, and the vertexing efficiency, $\varepsilon_{\textrm{vertex}}^{\mu\mu K}$, are found
to be equal to 99\% and are well reproduced by the MC simulation.
The reconstruction efficiency for hadrons
in the ID was verified in ref.~\cite{MinBias} for data and
simulation; for the kaons used in this analysis, the efficiency is obtained
from simulation.

The number of reconstructed $B^\pm$ mesons is obtained using a binned
maximum likelihood fit to the invariant mass distribution of the selected
candidates. The probability density function (pdf) for the signal is defined as
the sum of two Gaussians of relative fraction $f_1$ and corresponding widths
$\sigma_1$, $\sigma_2$, both centred at the reconstructed $B^{\pm}$ mass. The pdf for the background consists
of three components to model the following three sources of background:
\begin{itemize}
\item $B^\pm\to\Jpsi\,\pi^\pm$, where the kaon mass is wrongly assigned to
  the pion; this decay is Cabibbo suppressed with a relative
fraction of $4.9\%$~\cite{PDG2012} with respect to the signal decay;
it produces a resonant structure in the signal region that is modelled
with a Crystal Ball function (see appendix E of ref.~\cite{crystalball}.
\item $B^{\pm/0}\to\Jpsi\, K^{*\pm/0}\to\Jpsi \,(K\pi)^{\pm/0}$ and $B^{\pm/0}\to\Jpsi\,
  (K\pi)^{\pm/0}$, where the final-state pion is not associated to the decay vertex,
  creating a resonant structure displaced from the $B^\pm$ mass by
  about $m_\pi$, where $m_\pi$ is the mass of the pion; these partially reconstructed $B$-decays are modelled with a
  complementary error function.
\item Combinatorial background from random combinations of
  $\Jpsi$ (produced promptly in $\pp$ collisions or in feed-down from
  $B$-decays) with a track; it is modelled with an exponential  
  function. The background from muon pairs not originating from
  $\Jpsi$ decays is negligible after the
  full $B^{\pm}$ candidate selection.
\end{itemize}

\begin{figure}[hbt!]
  \centering 
    \includegraphics[width=0.6\textwidth]{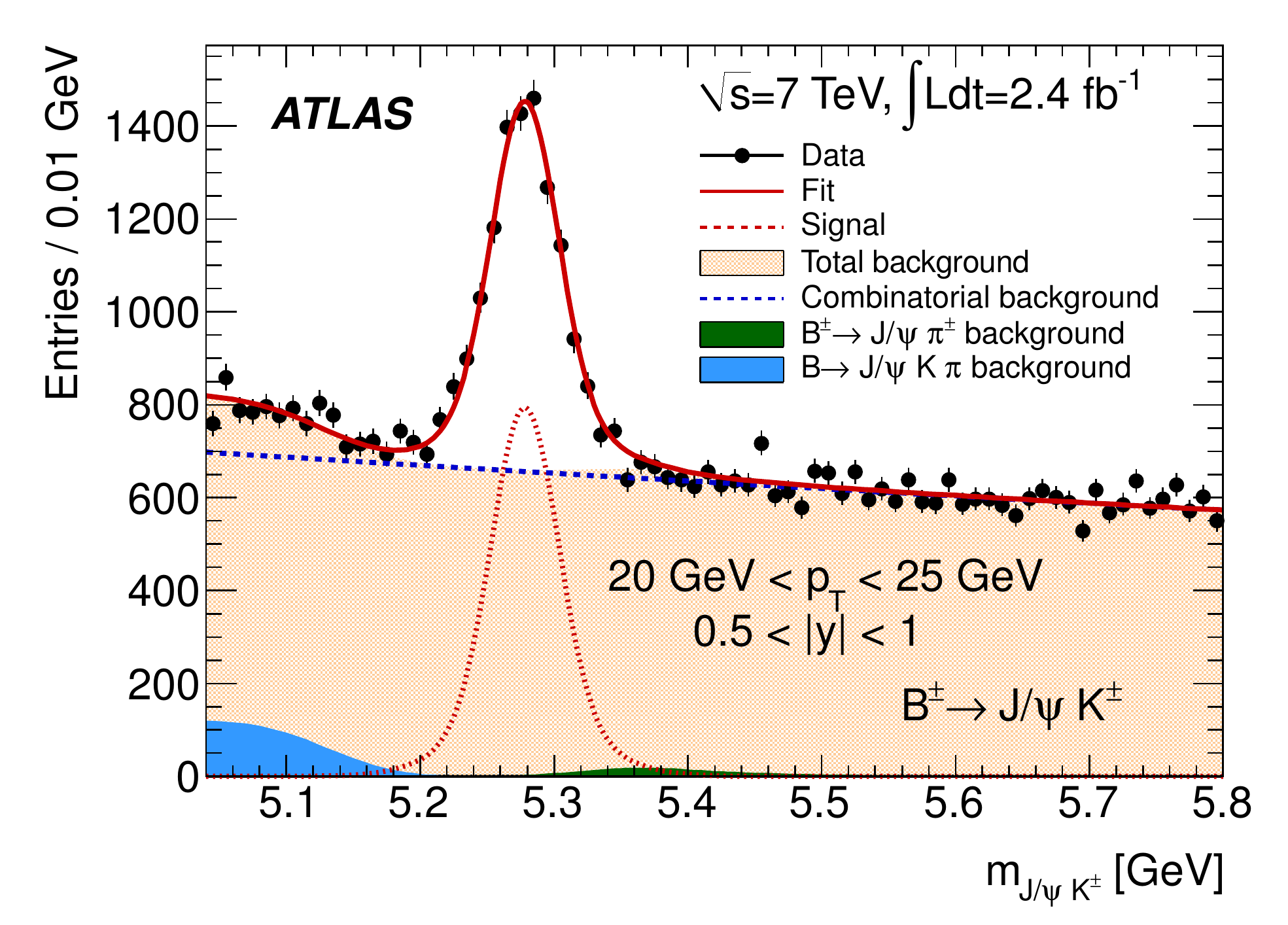}
    \caption{The observed invariant mass distribution of $B^\pm$
      candidates, $m_{\Jpsi K^\pm}$, with transverse momentum and
      rapidity in the range $20\GeV<\pt<25\GeV$, $0.5<|y|<1$ (dots), compared to
      the binned maximum likelihood fit (solid line). The error bars represent the statistical uncertainty. Also shown are the
      components of the fit as described in the legend.}
    \label{fig:fitEx}
\end{figure}

The extraction of the signal yield is done in two steps. First, the
shapes of the signal and the resonant background pdfs, which depend on the $\pt$ and
$y$ of the $B^\pm$ meson candidate, are obtained by fitting the
invariant mass distribution of signal and background events from MC
samples in each $(\pt, y)$ interval. 
Then the invariant mass distribution of the data is fitted in the same $(\pt, y)$ interval.
The parameters for the shape of the signal pdf ($\sigma_1$, $\sigma_2$ and $f_1$) and the
resonant backgrounds are fixed to the results of the fits to MC event samples. The relative normalisation of the
$B^\pm\to\Jpsi\,\pi^\pm$ decay to the signal is fixed to the
fraction of the world-average values for their branching ratios, and
is corrected for the difference in acceptance for
the two decay modes.
The reconstructed mass $m_{B^\pm}$ is obtained from data for the full
$\pt$ range in a rapidity interval by fitting the invariant mass distribution of the selected
candidates, and is fixed throughout the fits in $\pt$ intervals.
Therefore, when fitting the data in each $(\pt, y)$ interval, the free
parameters are the normalisation of the signal, the normalisation of
the partially reconstructed $B$-decays, and the slope of the combinatorial
background. The results of the fits to the invariant mass
distributions of the selected $B^\pm$ candidates from data are exemplified in figure~\ref{fig:fitEx} for an
interval of intermediate $\pt$ and central rapidity. The stability of the fit was tested with simulated samples of signal and background with statistical size similar to our data and no evidence of bias in the fit was found.

The total number of signal $B^\pm$ events observed in data in the full
$\pt$ and $y$ range covered by the analysis, $9\GeV<\pt<120\GeV$ and
$|y|<2.25$, before acceptance and efficiency corrections, is about $125600$.
These events populate four intervals in $|y|$ and eight intervals in
$\pt$ for the differential cross-section measurement.
The acceptance correction, $A$, has a small dependence on $y$ and
ranges from 4\% to 85\% from the low to the high $\pt$ intervals.
The efficiency $\varepsilon^{B^{+}}$ has a dependence on both $y$ and $\pt$ and
ranges from 25\% to 40\%. The relative difference between the efficiencies for reconstructing $B^+$ and $B^-$
mesons, $(\varepsilon^{B^{+}} - \varepsilon^{B^{-}}
)/\varepsilon^{B^{+}}$, has a
dependence on $\pt$ and ranges from 5\% to 2\%.

The assumption of equal $B^+$ and $B^-$ production is tested by
fitting the invariant mass distribution of $B^+$ and $B^-$ candidates
separately. The resulting yields, before applying efficiency
corrections, are $63530 \pm 840$ and $62090 \pm 840$ respectively,
where the quoted uncertainties are statistical.
Taking into account the different efficiencies for reconstructing $B^+$ and $B^-$ mesons,
the ratio of ${B^+}/ B^-$ is found to be consistent with unity, within the statistical
precision of this test.

\section{Systematic uncertainties}

Various sources of systematic uncertainty on the measurement of the
$B^+$ production cross-section are considered and discussed below:
\begin{enumerate}
\item \label{isyst:trigger}
{\bf Trigger.}
The trigger efficiency is obtained from
data in bins of $\pt$ and $q \cdot \eta$ of the muon, where $q$ is the
muon charge, using a tag-and-probe method~\cite{Upsilon}.
Then, the correction weights for the trigger efficiency
$w_{\textrm{trigger}}^{\Jpsi}$ (see eq.\;\eqref{eq:weights}) are
obtained from the fraction of the measured efficiency from data over the expectation from simulation in each $(\pt, q\cdot\eta)$ bin.
As the statistical components of the
uncertainty associated with the weights for the trigger efficiency are dominant, the
uncertainties on the cross-section are derived from a series of
pseudo-experiments by allowing the weights to fluctuate randomly
under a Gaussian assumption, according to their
assigned uncertainty.

\item \label{isyst:fit}
{\bf Fit.}
For the fit method, three sources of systematic uncertainty are
identified and considered to be uncorrelated. These are the shape of the
signal pdf, the reconstructed $B^\pm$ mass and the shape of the
background pdf. Below, the procedure to estimate the systematic
uncertainty from each source is described, and the resulting
uncertainties are added quadratically to obtain the total systematic
uncertainty from the fit method in each $(\pt, y)$ interval.
\begin{enumerate}
\item {\it Uncertainty on shape of the signal pdf.} This uncertainty
  is estimated with variations of the fit model, where the values of the signal pdf parameters $\sigma_1$,
  $\sigma_2$, $f_1$ are varied independently within their
  uncertainties derived from the fit to signal events from MC simulation.
  From these variations of the fit model, the largest absolute value
  of the signal yield variation is taken as the
  systematic uncertainty from the signal pdf shape, in order to
  account for the large correlations of these parameters. Two
  alternative pdfs were considered (three Gaussians, two Crystal Ball
  + Gaussian) and no significant differences in the signal yield were
  observed.  
  Among the various sources of systematic uncertainty considered for
  the fit method, the signal pdf is dominant and its contribution
  ranges from 1\% to 8\%.
\item {\it Uncertainty on the $B^\pm$ mass value.}
  The reconstructed mass $m_{B^\pm}$ is obtained from data by fitting
  the invariant mass distribution of all candidates
  with $\pt>9$\;GeV in each of the four rapidity intervals. The
  resulting values are used to fix this parameter when performing the
  fits in the various $(\pt, y)$ intervals and their statistical uncertainties (0.4--1.0\;MeV) are used to
  estimate the systematic uncertainty on the signal yield.
  The fits in the various $(\pt, y)$ intervals are repeated varying the 
  value of $m_{B^\pm}$ within its statistical uncertainty. The observed difference in the signal yield
  is smaller than 1\%.
\item {\it Uncertainty on the shape of the background pdf.} 
  The fit includes three components for the description of the
  background (see below), and each contributes as
  a possible source of uncertainty.
  In order to account for the large correlations between the three components, the
  systematic uncertainty assigned to the background modelling for each $(\pt, y)$
  interval is obtained after varying each component independently, and taking the largest observed difference in the signal yield.
  \begin{enumerate}
  \item Combinatorial background:
    with a polynomial instead of exponential shape for the
    combinatorial bacground; the observed relative difference in the signal
    yield ranges from 0.1\% to 4\%, where the larger change is
    observed for higher values of $y$ and $\pt$.    
  \item $B^\pm\rightarrow\Jpsi\,\pi^\pm$:
    for the resonant background from $B^\pm\rightarrow\Jpsi\,\pi^\pm$,
    the dominant uncertainty comes from the relative branching
    fraction of this decay with respect to the signal,
    which has an uncertainty of 10\%~\cite{PDG2012}. Varying this fraction in the fit within its
    uncertainty was found to have a small effect on the signal yield
    ($\sim$1\%).
  \item Partially reconstructed $B$-decays:
    the resonant background from partially reconstructed
    $B$-decays is modelled with a complementary error
    function. When varying its parameters within their
    uncertainties from the fits to background events from MC simulation, the observed difference in
    the signal yield is smaller than 1\%.
  \end{enumerate}
\end{enumerate}

\item {\bf Kaon track reconstruction.} \label{isyst:kaon}
  The efficiency of hadron reconstruction is determined from
  MC simulation and validated with data~\cite{MinBias}, with the uncertainty dominated by the material description.
  The uncertainty ranges with increasing rapidity from
  2\% to 4\% for the kaons used in this analysis.
  
\item \label{isyst:acc}
  {\bf Acceptance.} The acceptance in each $(\pt, y)$ interval has a relative uncertainty ranging from 1\% to 4\%,
  due to the size of the MC sample, which is assigned as systematic uncertainty. 

\item \label{isyst:muon}
  {\bf Muon reconstruction.}
  The muon reconstruction efficiency is obtained from
  data in bins of $\pt$ and $q \cdot \eta$ of the muon, using a
  tag-and-probe method~\cite{Upsilon}. 
  Then, the correction weights $w_{\textrm{MS}}^{\mu}$ (see eq.\;\eqref{eq:weights}) are
  obtained from the fraction of the measured efficiency from data over
  the expectation from simulation in each $(\pt, q\cdot\eta)$ bin.
  The uncertainties associated with the weights for the muon reconstruction
  efficiency are mainly
  statistical, so the same procedure as for the trigger
  efficiency is used to estimate the systematic uncertainty on the cross-section.
  In addition, there is also
  an uncertainty coming from the efficiency for
  reconstructing a muon in the ID with the selection criteria used in this
  analysis. This efficiency is found to be 99\% 
  with a systematic uncertainty of $0.5\%$ for each muon.
  
\item \label{isyst:vtx}
  {\bf {\boldmath $B^\pm$} vertex-finding efficiency.} 
  The vertex quality requirement has an efficiency of $\sim$99\% and is fairly
  independent of $\pt$ and $y$. It was estimated with data by comparing
  the signal yields in four rapidity intervals before and after applying
  this requirement and is found to be consistent with the expectation
  from MC simulation. A systematic uncertainty of $2\%$ is assigned to
  the cross-section, as the maximum difference observed between the estimate from data and
  the expectation from MC simulation for this efficiency.

\item \label{isyst:br} {\bf Branching ratio.}
  The total branching ratio of the selected decay, obtained by combining
  the branching ratios of the decays $B^\pm\to\Jpsi K^\pm$ and
  $\Jpsi\to\mumu$, has an uncertainty of 3.4\%~\cite{PDG2012}.

\item \label{isyst:lumi}
  {\bf Luminosity.}
  The luminosity calibration is based on data from van der Meer scans and has an uncertainty of 1.8\%~\cite{Lumi}.

\item \label{isyst:sigEff}
  {\bf Signal efficiency.}
  The efficiency correction factor for $B^\pm$ signal events is
  obtained from MC simulation (eq.\;\eqref{eq:eff}). The systematic uncertainty
  assigned to this factor has two components, which are added in quadrature:
  \begin{enumerate}
  \item {\it Uncertainty from the size of the MC sample.}  The sample used for the estimation of the efficiency correction factor corresponds to a luminosity similar to that of the data
    sample. Due to the size of this sample, the
    efficiency estimation has an uncertainty that is small ($\sim$1\%)
    in most intervals and becomes significant in the high-$\pt$
    interval 70--120\;GeV ($\sim$10\%). It is added quadratically to the rest of the sources
    of uncertainty.
  \item {\it Uncertainty from $K^+/K^-$ efficiency asymmetry.} The
    efficiencies for reconstructing $K^+$ and $K^-$ mesons are obtained
    from simulation and their relative difference is found to be $\sim$3.5\%.
    This difference is verified with data and the statistical
    uncertainty of the estimate from data is used to assign a
    systematic uncertainty of 1\%, which propagates to the
    cross-section through the sum of efficiencies ($\varepsilon^{B^+}
    + \varepsilon^{B^-}$) in eq.\;\eqref{eq:NBplus}.
  \end{enumerate}
\end{enumerate}

\begin{figure}
  \centering 
    \includegraphics[width=0.4\textwidth]{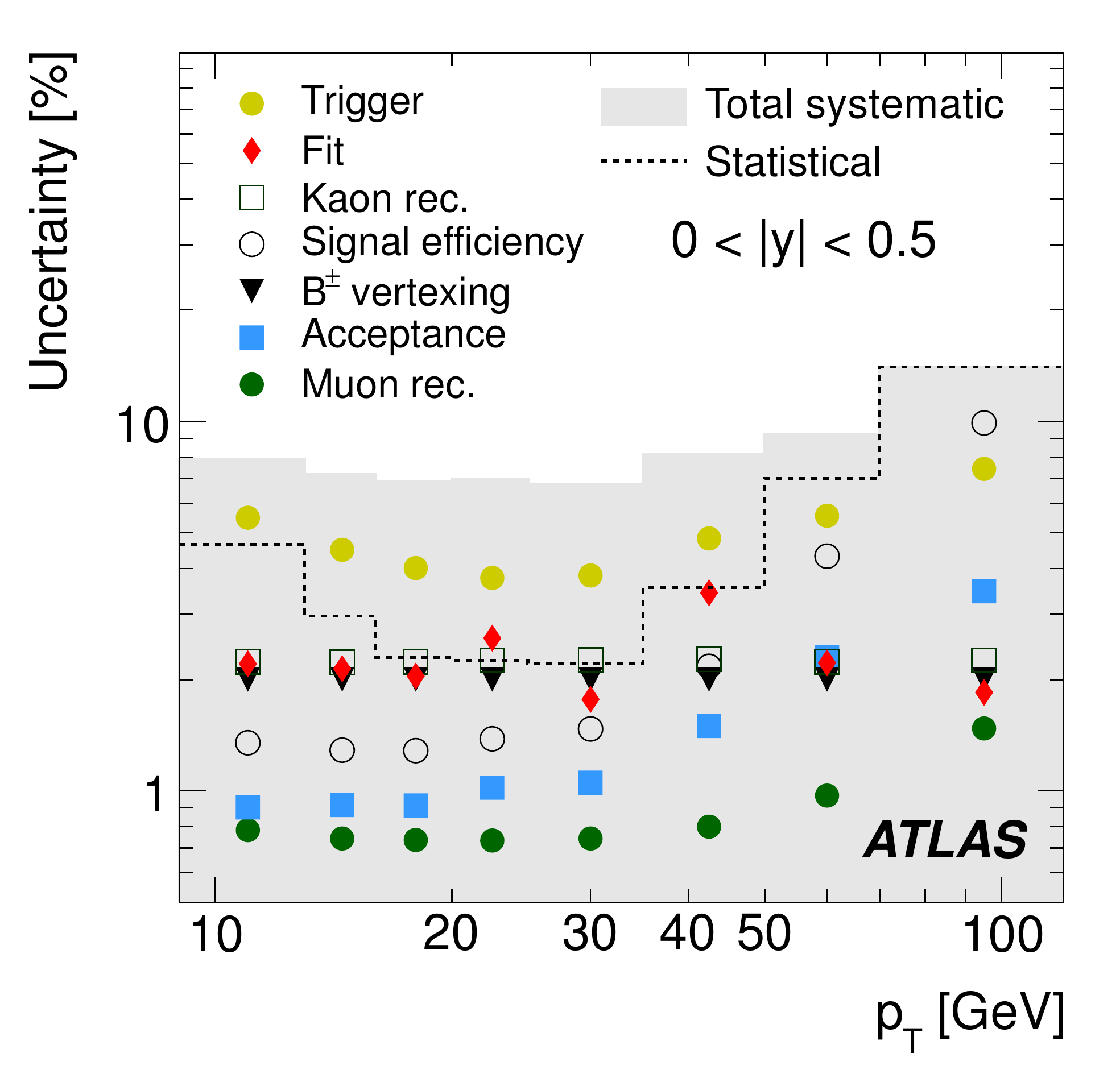}
    \includegraphics[width=0.4\textwidth]{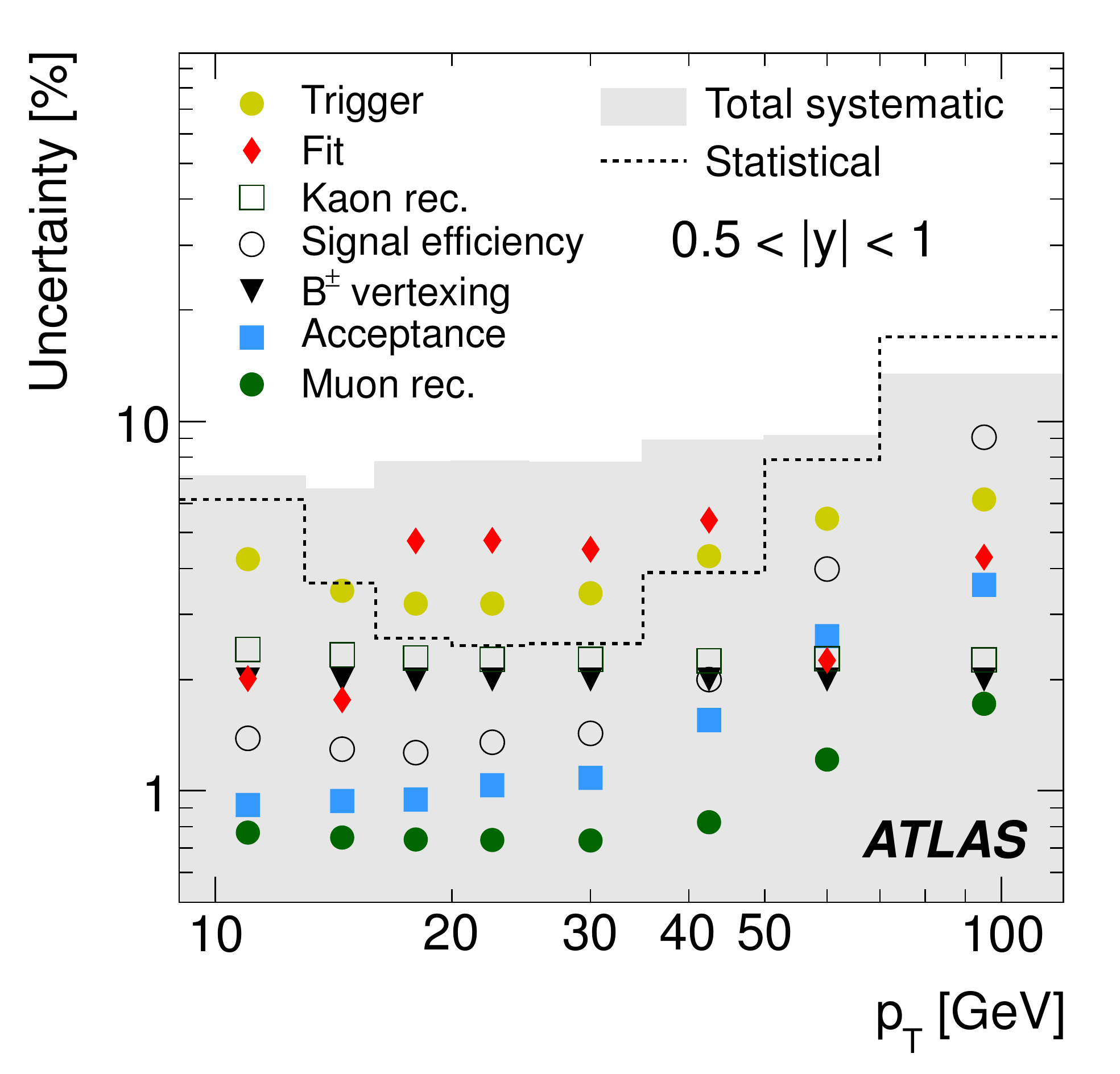}\\
    \includegraphics[width=0.4\textwidth]{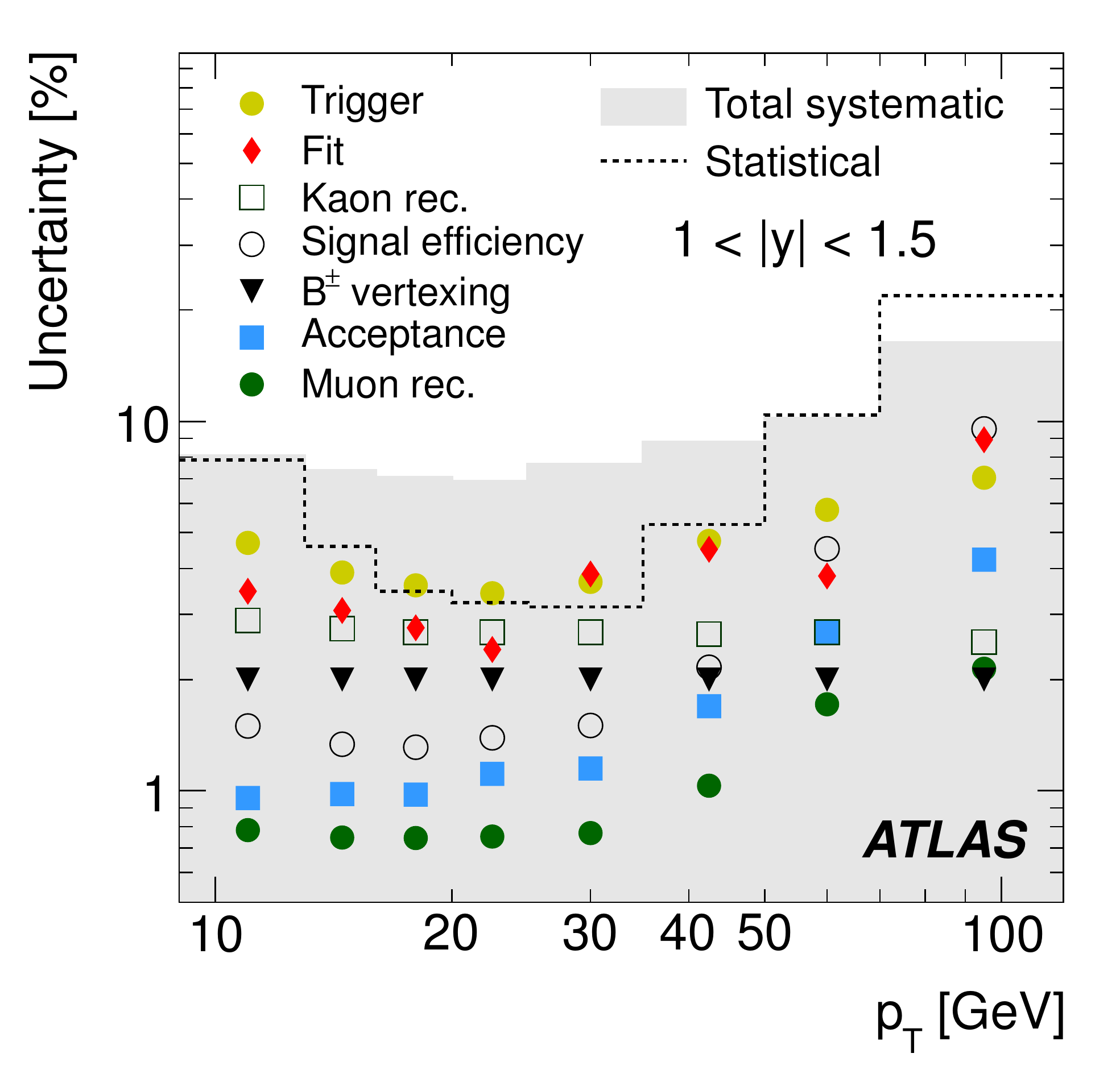}
    \includegraphics[width=0.4\textwidth]{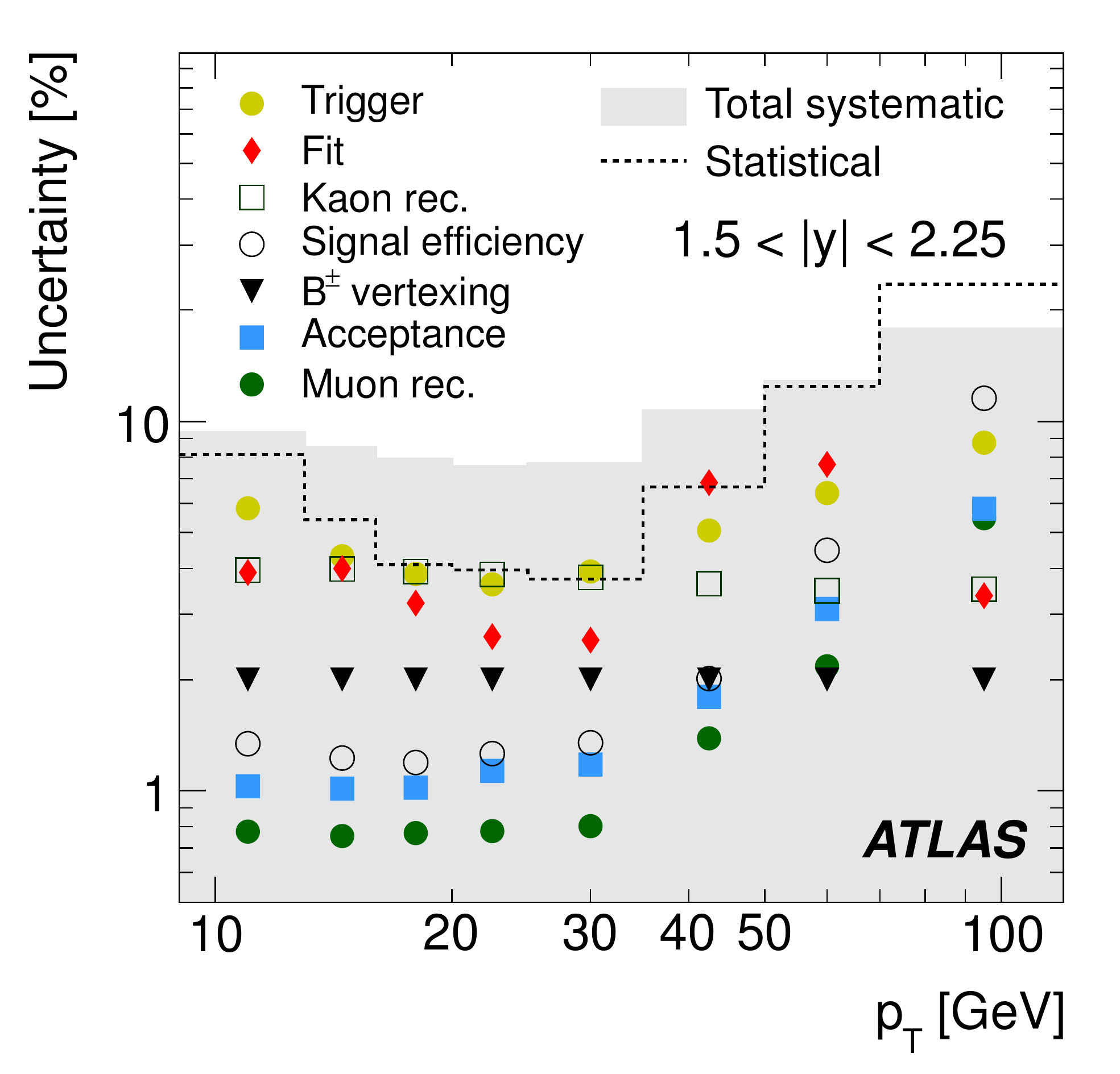}\\
    \caption{Relative systematic uncertainties on the cross-section determination
      as a function of $\pt$ for different rapidity ranges. The total systematic uncertainty (solid area),
      including uncertainties from luminosity (1.8\%) and branching
      ratio (3.4\%), is compared to the statistical uncertainty
      (dashed line).}
    \label{fig:eff_systematics}
\end{figure}

The range of these uncertainties is summarised in
table~\ref{tab:systematics}. Their breakdown in $(\pt, y)$ intervals
is given in figure\;\ref{fig:eff_systematics}. In the same figure, the total systematic uncertainty,
including the uncertainties from the luminosity and branching ratio, is compared to the
statistical precision of the measurement. In most intervals, the
systematic uncertainty dominates.
\begin{table}[b!]
  \centering 
    \begin{tabular}{lcccc}
      \hline\hline
      & \multicolumn{4}{c}{Relative uncertainty [\%]} \\
      Source & $|y| < 0.5$ & $0.5 < |y| <1$ &  $1< |y| < 1.5$ & $1.5 < |y| < 2.25$\\
      \hline
      Statistical uncertainty & 2.2--14 & 2.5--17 & 3.2--22 & 3.8--24 \\
      Total systematic uncertainty & 6.7--14 & 6.5--13 & 6.9--16 & 7.6--18\\
      \hline
      \ref{isyst:trigger}. Trigger & 3.8--7.4 & 3.2--6.2 & 3.4--7.0 & 3.6--8.8\\
      \ref{isyst:fit}. Invariant mass fit & 1.8--3.4 & 1.7--5.3 & 2.4--8.9 & 2.6--7.6\\
      \ref{isyst:kaon}. Kaon reconstruction & 2.2 & 2.2--2.4 & 2.5--2.9 & 3.5--4.0\\
      \ref{isyst:acc}. Acceptance & 0.9--3.5 & 0.9--3.6 & 1.0--4.2 & 1.0--5.8 \\
      \ref{isyst:muon}. Muon reconstruction & 0.5--1.3 & 0.5--1.7 & 0.5--2.1 & 0.6--5.4\\
      \ref{isyst:vtx}. $B^\pm$ vertexing & 2.0 & 2.0 & 2.0 & 2.0 \\
      \ref{isyst:br}. Branching ratio  & 3.4 & 3.4 & 3.4 & 3.4 \\
      \ref{isyst:lumi}. Luminosity & 1.8 & 1.8 & 1.8 & 1.8\\      
      \ref{isyst:sigEff}. Signal efficiency & 1.3--10 & 1.3--9.1 & 1.3--9.5  & 1.2--12\\  
      \hline\hline
    \end{tabular}
    \caption{The statistical and total systematic uncertainties on the
      cross-section measurement in different ranges of rapidity $y$.
      The contributions from the various sources of systematic
      uncertainty are also given. The range of values quoted for some
      of the uncertainties represent the lower and upper limit of the
      uncertainty over the $\pt$ range in a given rapidity range.}
  \label{tab:systematics}
\end{table}

Additional sources of systematic uncertainty were examined, but were
found to be less significant and were neglected.
Residual effects related to final-state radiation have been determined
to be smaller than 1\% and are neglected.
Differences in the underlying kinematic
distributions modelled by the \Pythia~and NLO generators, including
parton distribution functions, were
considered. The impact on the acceptance and the signal efficiency was estimated by reweighting
the kinematic distributions of \Pythia~to those of \POWHEG~and
\MCatNLO.
The largest effect is seen in the high-rapidity intervals ($1.5<|y|<2.25$), where
the maximum relative difference observed is 1\%, with a
statistical uncertainty of the same order, while in most $(\pt, y)$ intervals
the effect is very small ($\sim$0.1\%).
Bin-to-bin migration of signal events due to finite detector resolution is
studied with MC simulation.
It is found to be a small effect ($<$0.5\%), which is
included in the definition of signal efficiency (eq.\;\eqref{eq:eff}).
Potential effects in the calculation of the signal efficiency
due to the difference between the momentum scales in data and
MC simulation are expected to be larger in the lower $\pt$ intervals used in
this analysis, where they were estimated to be smaller than 0.5\%.

\section{Cross-section results} 
\label{sec:results}
Using eq.\;\eqref{eq:xsec}, the differential cross-section
for $B^+$ production times the product of branching ratios $\mathcal{B}$ is obtained as a function of $\pt$ and $y$ of the $B^+$
meson and the results are shown in tables\;\ref{tab:xsRap12} and
\ref{tab:xsRap34}, averaged over each $(\pt,y)$ interval.
The double-differential cross-section is integrated over $\pt$ to obtain the differential cross-section $\textrm{d}\sigma/\textrm{d}y$, or over rapidity to obtain
$\textrm{d}\sigma/\textrm{d}\pt$, and results are reported in
tables\;\ref{tab:xs_dpt} and \ref{tab:xs_dy}.
When summing over the intervals in $\pt$ or rapidity,
the systematic uncertainty from each source is calculated from
the linear sum of the contributions from each
interval, as they are correlated.
Tabulated results of the measurements presented in this paper are
available in \textsc{HepData}~\cite{hepdata}.
\begin{table}[h!]
  \centering 
    \begin{tabular}{l|r@{$\pm$}c@{$\pm$}l|r@{$\pm$}c@{$\pm$}l}  
      \hline\hline
      $\pt$ interval & \multicolumn{6}{|c}{$\frac{\textrm{d}^2\sigma}{\textrm{d}\pt \textrm{d}y}\cdot
        {\pazocal B} (B^+ \to\Jpsi K^+) \cdot {\pazocal B} (\Jpsi\to\mu^+ \mu^- )$ [pb/GeV]} \\
      ~[GeV] & \multicolumn{3}{c}{$0 < |y| < 0.5$} & \multicolumn{3}{c}{$0.5 < |y| < 1$}\\
      \hline     
      ~~9--13 & 24.5 \,&\, 1.1 \,&\, 1.7 & 21.7 \,&\, 1.3 \,&\, 1.4 \\
      13--16 & 8.7 \,&\, 0.3 \,&\, 0.6 & 8.5 \,&\, 0.3 \,&\, 0.5 \\
      16--20 & 3.76 \,&\, 0.09 \,&\, 0.22 & 3.9 \,&\, 0.10 \,&\, 0.27 \\
      20--25 & 1.54 \,&\, 0.04 \,&\, 0.09 & 1.57 \,&\, 0.04 \,&\, 0.11 \\
      25--35 & 0.467 \,&\, 0.010 \,&\, 0.027 & 0.468 \,&\, 0.012 \,&\, 0.033 \\
      35--50 & 0.097 \,&\, 0.003 \,&\, 0.007 & 0.095 \,&\, 0.004 \,&\, 0.008 \\
      50--70 & 0.0165 \,&\, 0.0012 \,&\, 0.0014 & 0.0178 \,&\, 0.0014 \,&\, 0.0015 \\
      70--120 & 0.00188 \,&\, 0.00026 \,&\, 0.00025 & 0.00202 \,&\, 0.00034  \,&\, 0.00026 \\
      \hline\hline
    \end{tabular}
    \caption{Differential cross-section measurement for $B^+$
      production multiplied by the branching ratio to the final state,
      averaged over each $(\pt,y)$ interval in the rapidity range
      $|y|<0.5$ and $0.5<|y|<1$.
      The first quoted uncertainty is statistical, the second
      uncertainty is systematic.}
    \label{tab:xsRap12}
\end{table}
\begin{table}[h!]
  \centering 
  \begin{tabular}{l|r@{$\pm$}c@{$\pm$}l|r@{$\pm$}c@{$\pm$}l}
      \hline\hline
      $\pt$ interval & \multicolumn{6}{|c}{$\frac{\textrm{d}^2\sigma}{\textrm{d}\pt \textrm{d}y}\cdot
        {\pazocal B} (B^+ \to\Jpsi K^+) \cdot {\pazocal B} (\Jpsi\to\mu^+ \mu^- )$ [pb/GeV]} \\
      ~[GeV] & \multicolumn{3}{c}{$1 < |y| < 1.5$} & \multicolumn{3}{c}{$1.5 < |y| < 2.25$}\\
      \hline
      ~~9--13 & 23.6 \,&\, 1.9 \,&\, 1.7 & 22.3 \,&\, 1.8 \,&\, 1.9 \\
      13--16 & 8.0 \,&\, 0.4 \,&\, 0.5 & 7.1 \,&\, 0.4 \,&\, 0.6 \\
      16--20 & 3.29 \,&\, 0.11 \,&\, 0.20 & 2.90 \,&\, 0.12 \,&\, 0.21 \\ 
      20--25 & 1.32 \,&\, 0.04 \,&\, 0.08 & 1.08 \,&\, 0.04 \,&\, 0.07 \\
      25--35 & 0.408 \,&\, 0.013 \,&\, 0.028 & 0.312 \,&\, 0.012 \,&\, 0.022  \\
      35--50 & 0.073 \,&\, 0.004 \,&\, 0.006 & 0.055 \,&\, 0.004 \,&\, 0.006 \\
      50--70 & 0.0135 \,&\, 0.0014 \,&\, 0.0013 & 0.0097 \,&\, 0.0012 \,&\, 0.0012 \\
      70--120 & 0.00095 \,&\, 0.00021 \,&\, 0.00015 & 0.00083 \,&\, 0.00019 \,&\, 0.00014 \\
      \hline\hline
    \end{tabular}
    \caption{Differential cross-section measurement for $B^+$ production
      multiplied by the branching ratio to the final state, averaged
      over each $(\pt,y)$ interval in the rapidity range $1<|y|<1.5$ and
      $1.5<|y|<2.25$. The first quoted uncertainty is statistical, the
      second uncertainty is systematic.}
    \label{tab:xsRap34}
\end{table}
\begin{table}[h!]
  \centering 
    \begin{tabular}{l|r@{$\pm$}c@{$\pm$}l}       
      \hline\hline
      $\pt$ interval & \multicolumn{3}{|c}{$\frac{\textrm{d}\sigma}{\textrm{d}\pt}\cdot
        {\pazocal B} (B^+ \to\Jpsi K^+) \cdot {\pazocal B} (\Jpsi\to\mu^+ \mu^- )$ [pb/GeV]} \\
      ~[GeV] & \multicolumn{3}{c}{$|y| < 2.25$}\\
      \hline
      ~~9--13 & ~~~~~~~~~~~~~~~~~103 \,&\, 4 \,&\, 8  \\
      13--16 & ~~~~~~~~~~~~~~~~~36.0 \,&\, 0.8 \,&\, 2.3  \\
      16--20 & ~~~~~~~~~~~~~~~~~15.3 \,&\, 0.3 \,&\, 1.0  \\
      20--25 &~~~~~~~~~~~~~~~~~ 6.1 \,&\, 0.1 \,&\, 0.4  \\
      25--35 & ~~~~~~~~~~~~~~~~~1.81 \,&\, 0.03 \,&\, 0.12  \\
      35--50 & ~~~~~~~~~~~~~~~~~0.348 \,&\, 0.008 \,&\, 0.028  \\
      50--70 &~~~~~~~~~~~~~~~~~ 0.062 \,&\, 0.003 \,&\, 0.005  \\
      70--120 &~~~~~~~~~~~~~~~~~ 0.0061 \,&\, 0.0006 \,&\, 0.0007  \\      
      \hline\hline
    \end{tabular}
    \caption{Differential cross-section measurement for $B^+$ production
      multiplied by the branching ratio to the final state, averaged
      over each $\pt$ interval in the rapidity range $|y|<2.25$.
      The first quoted uncertainty is statistical, the second
      uncertainty is systematic.}
    \label{tab:xs_dpt}      
\end{table}
\begin{table}[htbp!]
\centering 
    \begin{tabular}{r@{--}l|r@{$\pm$}c@{$\pm$}l}       
      \hline\hline
       \multicolumn{2}{c}{$|y|$ interval} &  \multicolumn{3}{|c}{$\frac{\textrm{d}\sigma}{\textrm{d}y}\cdot
        {\pazocal B} (B^+ \to\Jpsi K^+) \cdot {\pazocal B} (\Jpsi
        \to\mu^+ \mu^- )$ [pb]}\\
      \multicolumn{2}{c}{} &  \multicolumn{3}{|c}{9\;GeV\;$<\pt<120$\;GeV}\\      
      \hline
      0.0&0.5 & ~~~~~~~~~~~~~~~~~~~~~~~~154 \,&\, 5 \,&\, 10  \\
      0.5&1.0 & ~~~~~~~~~~~~~~~~~~~~~~~~143 \,&\, 6 \,&\, 9  \\
      1.0&1.5 & ~~~~~~~~~~~~~~~~~~~~~~~~144 \,&\, 8 \,&\, 10  \\
      1.5&2.25 & ~~~~~~~~~~~~~~~~~~~~~~~~132 \,&\, 7 \,&\, 11  \\
      \hline\hline
    \end{tabular}
    \caption{Differential cross-section measurement for $B^+$ production
      multiplied by the branching ratio to the final state, averaged
      over each $y$
      interval in the $\pt$ range 9\;GeV$<\pt<120$\;GeV.
      The first quoted uncertainty is statistical, the second uncertainty is
      systematic.}
    \label{tab:xs_dy}
\end{table}

Using the world-average values for the branching ratio $\cal{B}$, the
differential cross-sections obtained are compared to predictions of
\POWHEG~(+\Pythia) and \MCatNLO~(+\HERWIG) and the FONLL
approximations.
For \POWHEG~and
\MCatNLO~the CT10~\cite{CTEQ10} parameterisation for the parton distribution function
of the proton is used, while for FONLL calculations the
CTEQ6.6~\cite{CTEQ6} parameterisation is used.
In all cases, a $b$-quark mass of $4.75\pm 0.25$\;GeV is used, with the
renormalisation and factorisation
scales, $\mu_r$, $\mu_f$, set to $\mu_r=\mu_f=\mu$, where $\mu$ has different
definitions for the \POWHEG, \MCatNLO~and FONLL predictions.\footnote{For \POWHEG: $\mu^2 = m^2_Q + (m_{Q\bar{Q}}^2/4-m_{Q}^2) \sin^2\theta_Q ~$, where $m_{Q\bar{Q}}$ is the invariant mass of the
 $Q\bar{Q}$ system and $\theta_Q$ is the polar angle of the heavy quark
 in the $Q\bar{Q}$ rest frame. For \MCatNLO: $\mu^2 = m_Q^2 +
( p_{\textrm{T, }Q} + p_{\textrm{T, }\bar{Q}} )^2/4$, where $p_{\textrm{T, }Q}$ and
$p_{\textrm{T, }\bar{Q}}$ are the transverse momenta of the produced heavy
quark and antiquark respectively, and $m_Q$ is the heavy-quark
 mass. For FONLL: $\mu = \sqrt{m_Q^2 + p_{\textrm{T, }Q}^2}$.}
The predictions are quoted with uncertainties due to the
$b$-quark mass and renormalisation and factorisation
scales.
Uncertainties from factorisation and renormalisation scales are estimated by varying them
independently up and down by a factor of two~\cite{fonll2012}. 

\begin{figure}[h!]
  \centering 
    \includegraphics[width=0.49\textwidth]{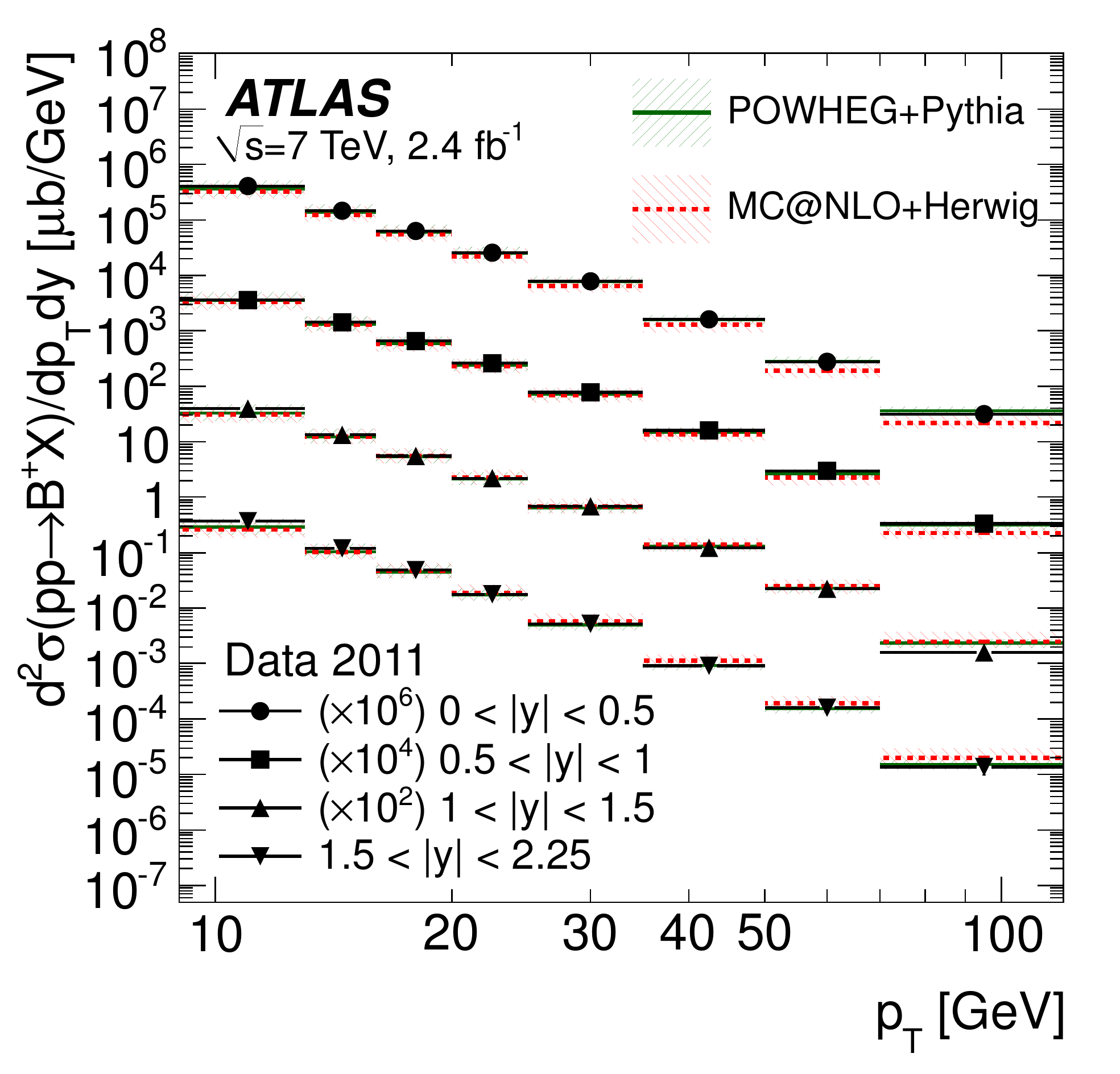}
    \caption{Double-differential cross-section of $B^+$ production as
      a function of $\pt$ and $y$, averaged over each $(\pt,y)$
      interval and quoted at its centre.
      The data points are compared to NLO predictions from \POWHEG~and
      \MCatNLO.
      The shaded areas around the theoretical predictions reflect the
      uncertainty from renormalisation and factorisation scales and
      the $b$-quark mass.}
    \label{fig:xs_dptdy}
\end{figure}

\begin{figure}[htb!]
 \centering 
    \includegraphics[width=0.45\textwidth]{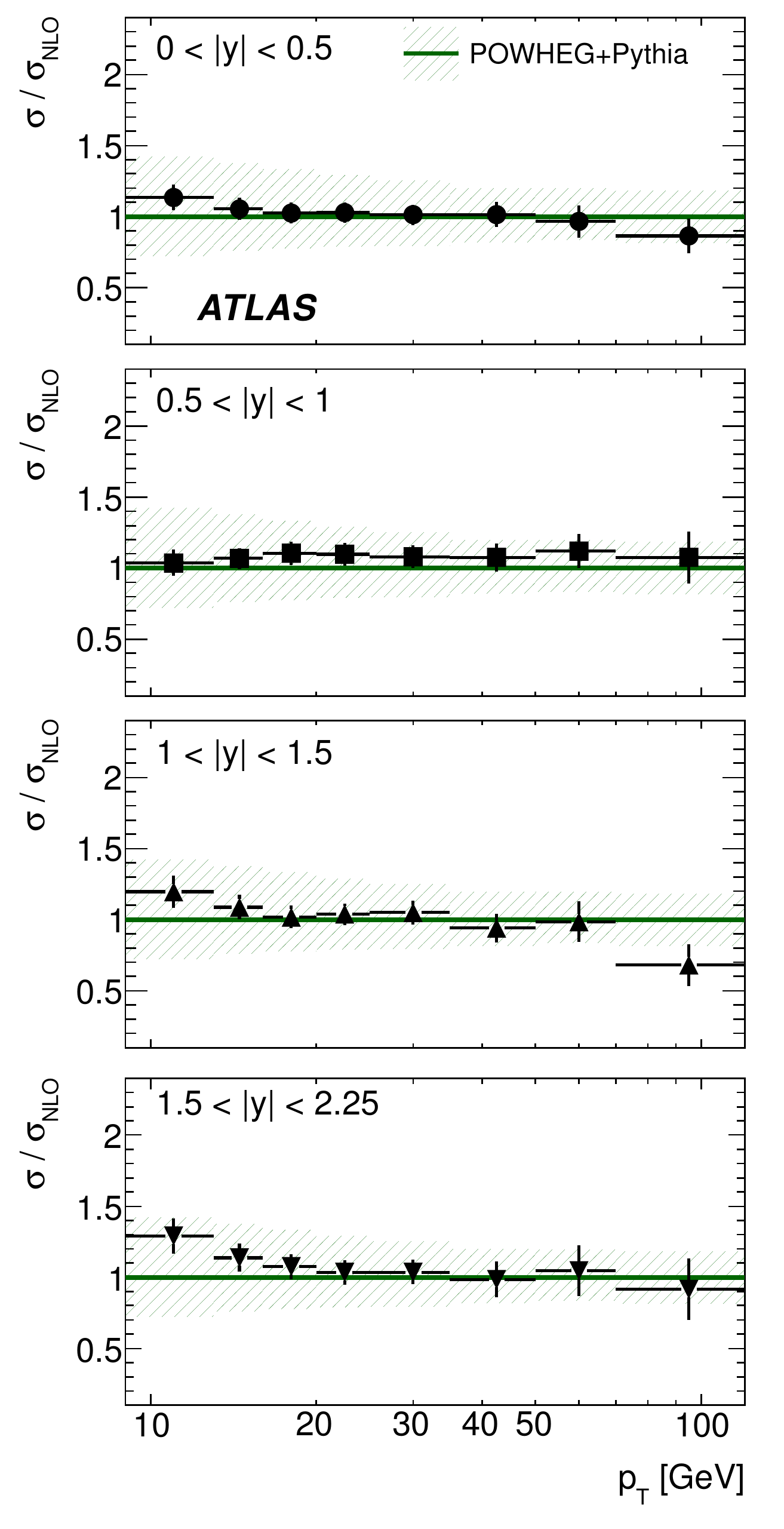}
    \includegraphics[width=0.45\textwidth]{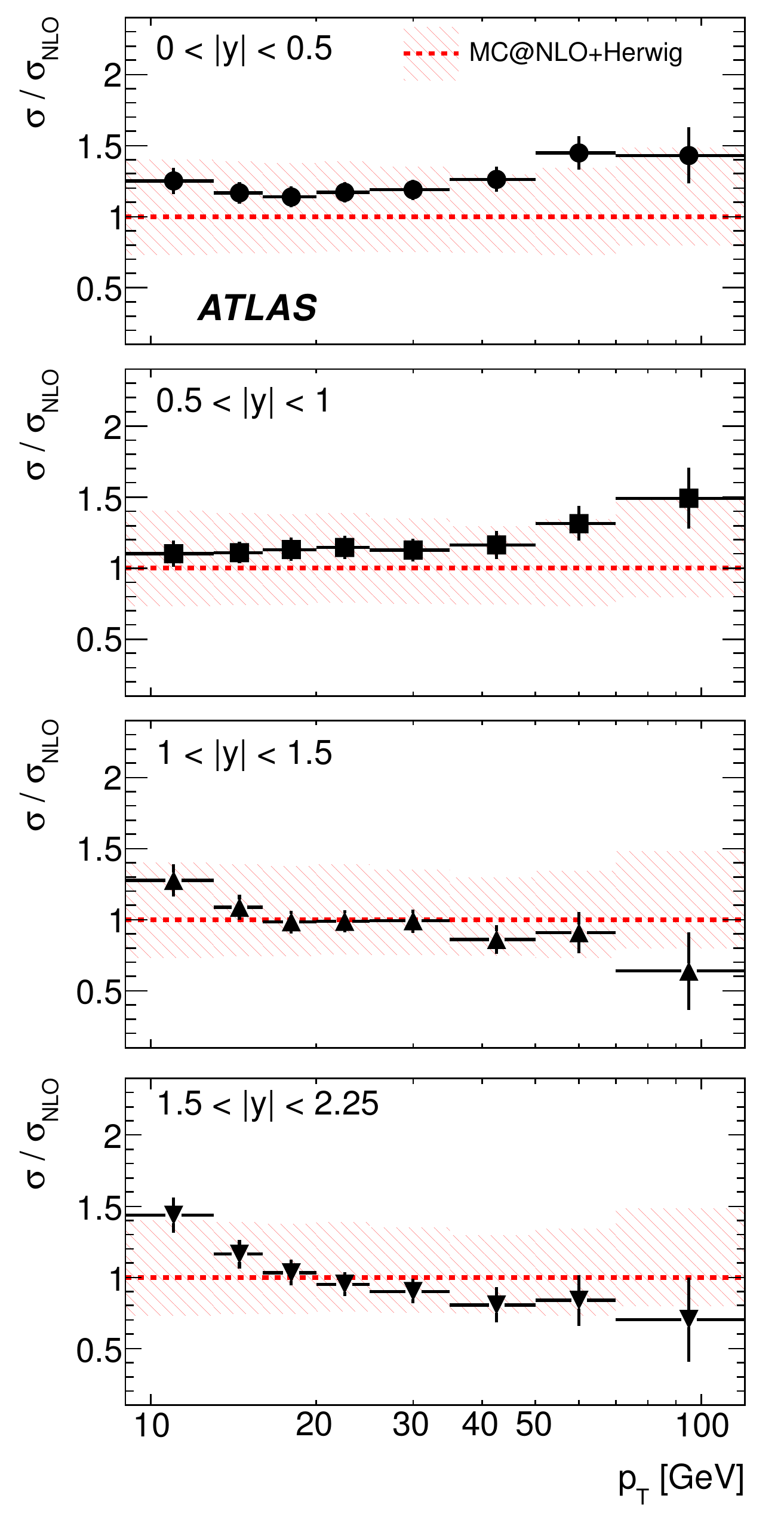}
    \caption{Ratio of the measured cross-section to the theoretical
      predictions ($\sigma/\sigma_{\textrm{NLO}}$) of \POWHEG~(left)
      and \MCatNLO~(right) in eight $\pt$ intervals in four rapidity
      ranges.
      The points with error bars correspond to data with their
      associated uncertainty, which is the combination of the
      statistical and systematic uncertainty.
      The shaded areas around the theoretical predictions reflect the
      uncertainty from renormalisation and factorisation scales and
      the $b$-quark mass.}
    \label{fig:xs_dptdy_ratioToMC}
\end{figure}

\POWHEG~and \MCatNLO~predictions are compared with the double-differential
cross-section measurement in figure\;\ref{fig:xs_dptdy}.
To allow a better comparison between the measured cross-sections and
the NLO predictions, figure\;\ref{fig:xs_dptdy_ratioToMC} shows their
ratio for each rapidity range separately for \POWHEG~and \MCatNLO.
The data are in good agreement with \POWHEG~in all rapidity intervals.
\MCatNLO, however, predicts a lower production cross-section at low
$\pt$ and a $\pt$ spectrum that is softer than the
data for $|y|<1$ and harder than the data for $|y|>1$.
In the integration of the four rapidity intervals, this effect
averages out and the prediction of the cross-section
$\textrm{d}\sigma/\textrm{d}\pt$ is compatible with data.

\begin{figure}[htbp!]
  \centering 
    \includegraphics[width=0.45\textwidth]{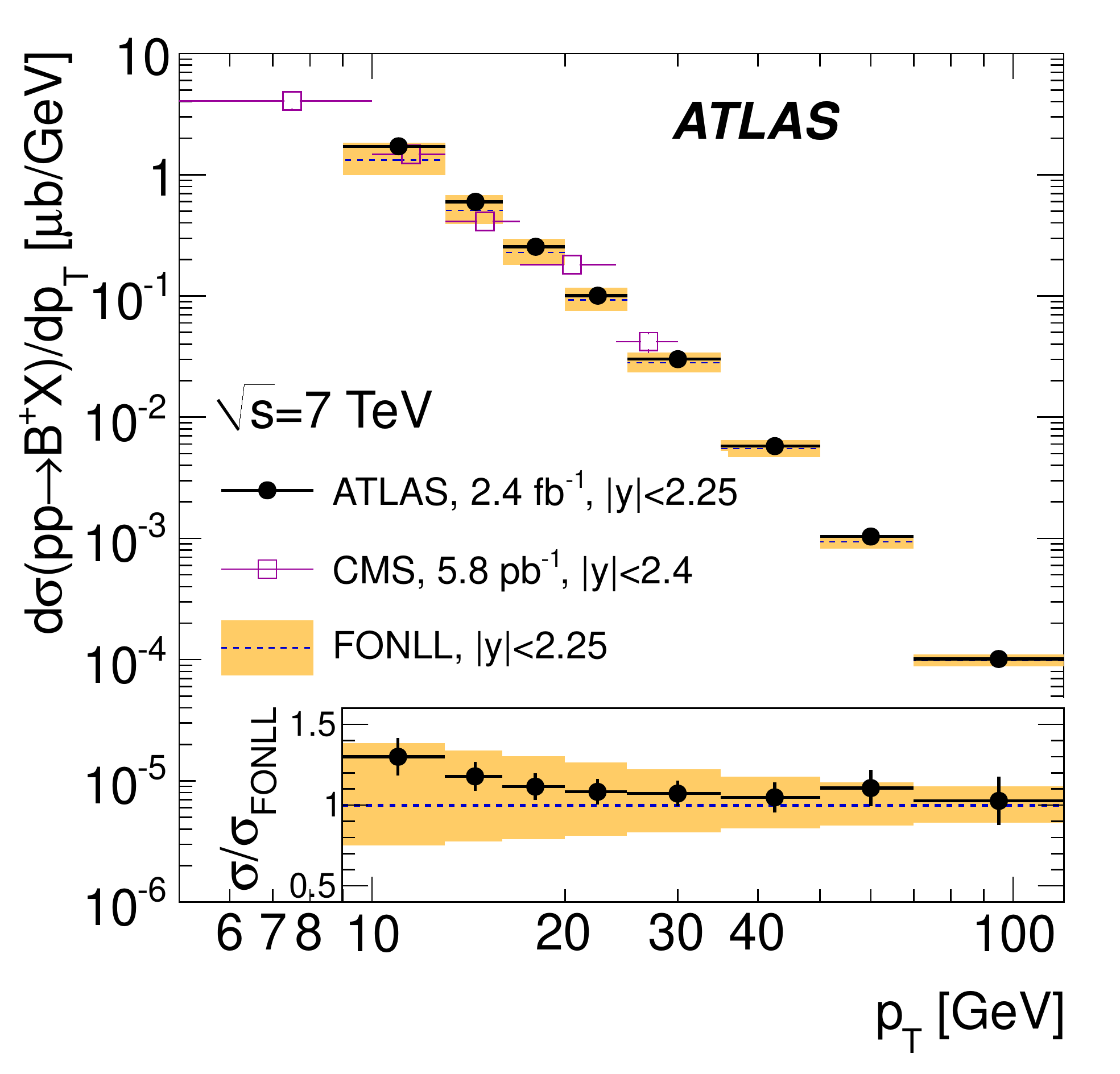}
    \caption{Differential cross-section of $B^+$ production vs $\pt$,
      integrated over rapidity.      
      The solid circle points with error bars correspond to the
      differential cross-section measurement of ATLAS with total
      uncertainty (statistical and systematic) in the rapidity range
      $|y|<2.25$, averaged over each $\pt$ interval and quoted at its
      centre.
      For comparison, data points from CMS are also shown, for a
      measurement covering $\pt<30$\;GeV and $|y|<2.4$~\cite{cms1}.
      Predictions of the FONLL calculation~\cite{fonll2012} for $b$-quark
      production are also compared with the data, assuming a hadronisation
      fraction of $f_{\bar{b}\to B^+}$ of $(40.1\pm 0.8)\%$~\cite{PDG2012} to
      fix the overall scale. Also shown is the ratio of the measured
      cross-section to the predictions of the FONLL calculation
      ($\sigma/\sigma_{\textrm{FONLL}}$). The upper and lower
      uncertainty limits on the prediction were obtained considering
      scale and $b$-quark mass variations.}
    \label{fig:xs_dpt}
\end{figure}

The FONLL prediction is compared with the measured differential cross-section
$\textrm{d}\sigma/\textrm{d}\pt$ in figure\;\ref{fig:xs_dpt}.
In this figure, the results from CMS~\cite{cms1} for $B^+$ meson production as a function
of $\pt$, covering the rapidity range $|y|<2.4$, are shown for
comparison.
The FONLL prediction is in good agreement with the data concerning the behaviour in rapidity
and $\pt$, within the theoretical uncertainties.

\begin{figure}[htbp!]
  \centering 
    \includegraphics[width=0.45\textwidth]{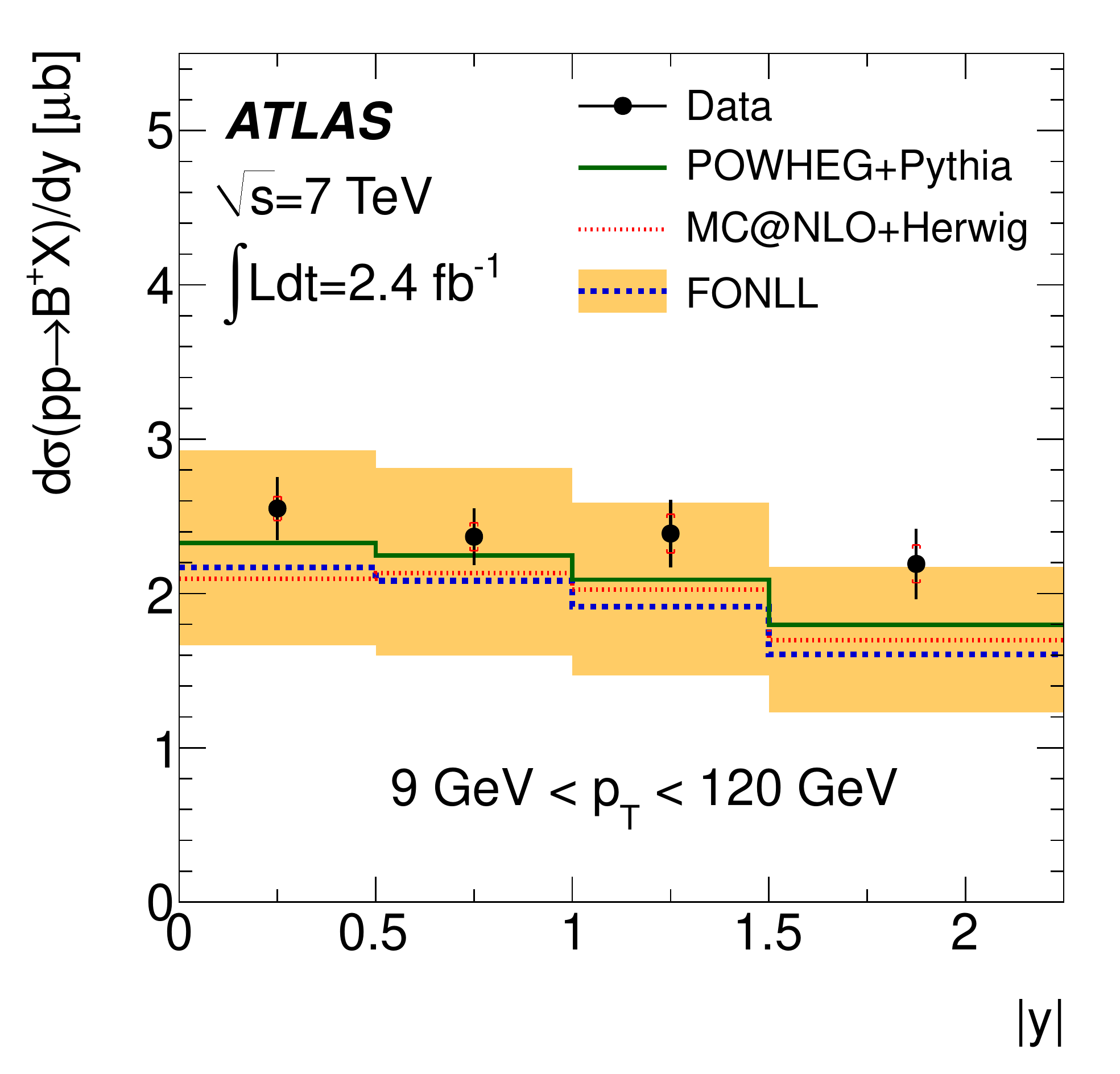}
    \caption{Differential cross-section of $B^+$ production vs rapidity,
      integrated over $\pt$.      
      Points with error bars correspond to the differential
      cross-section measurement
      with total uncertainty (lines on the error bars indicate the
      statistical component) in the $\pt$ range $9\GeV<\pt<120\GeV$,
      averaged over each rapidity interval and quoted at its centre.
      \POWHEG, \MCatNLO~and FONLL predictions are also given for
      comparison. The FONLL prediction is quoted with upper and lower
      uncertainty limits, which were obtained considering scale and
      $b$-quark mass variations.
      The relevant uncertainties of the predictions of \POWHEG~and
      \MCatNLO~are of the same order and are not shown.}
    \label{fig:xs_dy}
\end{figure}

All available predictions for $\textrm{d}\sigma/\textrm{d}y$ are
compared with data in figure\;\ref{fig:xs_dy}.
The measured cross-section has a small rapidity dependence and is in
agreement with the predictions within their uncertainties.
The theoretical uncertainties in all cases are large ($\sim$30\%) and are similar for the \POWHEG, \MCatNLO~and
FONLL predictions.

The integrated $B^+$ production cross-section in the kinematic range
$9\GeV<\pt<120\GeV$ and $|y|<2.25$ is:
\begin{equation*}
  \sigma(\pp\to B^+ X) =10.6 \pm 0.3~{\textrm{(stat.)}} \pm
  0.7 ~{\textrm {(syst.)}} \pm 0.2 ~{\textrm{(lumi.)}} \pm
  0.4~ {(\cal{B})} ~\upmu \textrm{b}.
\end{equation*}
The FONLL prediction, with its theoretical uncertainty from
the renormalisation and factorisation scale and the $b$-quark mass,
is:
\begin{equation*}
  \sigma(\pp\to b X) \cdot f_{\bar{b}\to B^+} =8.6^{+3.0}_{-1.9}~ \textrm{(scale)} \pm
  0.6 ~(m_b) ~\upmu \textrm{b} ~,
\end{equation*}
where $f_{\bar{b}\to B^+} = (40.1\pm 0.8)\%$~\cite{PDG2012} is the
world-average value for the hadronisation fraction. The corresponding predictions
of \POWHEG~and \MCatNLO~are $9.4~\upmu$b and
$8.8~\upmu$b, respectively, with theoretical uncertainties similar to
those of the FONLL prediction.

\section{Conclusions}
\label{section:conclusion}

The differential cross-section for $B^+$ meson production has been
studied with 2.4\;fb$^{-1}$ of $\pp$ collision data at
$\sqrt{s}=7$\;TeV, recorded in 2011 with the ATLAS detector at the LHC.
The cross-section was measured as a function of transverse momentum and rapidity in
the range $9\GeV<\pt<120\GeV$ and $|y|<2.25$, and quoted with a total uncertainty of 7\%--30\%
with the main source of uncertainty being systematic.
The next-to-leading-order QCD calculation is compatible with the measured differential cross-section.
The predictions are obtained within the \POWHEG~and \MCatNLO~frameworks and are quoted with an uncertainty
from renormalisation and factorisation scales and $b$-quark mass of
the order of 20\%--40\%. 
Within these uncertainties, \POWHEG+\Pythia~is in agreement with the
measured integrated cross-sections and with the dependence on $\pt$ and $y$.
At low $|y|$, \MCatNLO+\HERWIG~predicts a lower production cross-section
and a softer $\pt$ spectrum than the one observed in data, while for
$|y| > 1$ the predicted $\pt$ spectrum becomes harder than observed in
data.
The FONLL calculation for $\sigma(\pp\to b\,X)$ is compared to the data, assuming a hadronisation
fraction $f_{\bar{b}\to B^+}$ of $(40.1\pm 0.8)\%$~\cite{PDG2012},
and is in good agreement with the measured differential cross-section
$\textrm{d}\sigma/\textrm{d}\pt$, within the theoretical uncertainty.

\acknowledgments 
We thank CERN for the very successful operation of the LHC, as well as the
support staff from our institutions without whom ATLAS could not be
operated efficiently.

We acknowledge the support of ANPCyT, Argentina; YerPhI, Armenia; ARC,
Australia; BMWF and FWF, Austria; ANAS, Azerbaijan; SSTC, Belarus; CNPq and FAPESP,
Brazil; NSERC, NRC and CFI, Canada; CERN; CONICYT, Chile; CAS, MOST and NSFC,
China; COLCIENCIAS, Colombia; MSMT CR, MPO CR and VSC CR, Czech Republic;
DNRF, DNSRC and Lundbeck Foundation, Denmark; EPLANET, ERC and NSRF, European Union;
IN2P3-CNRS, CEA-DSM/IRFU, France; GNSF, Georgia; BMBF, DFG, HGF, MPG and AvH
Foundation, Germany; GSRT and NSRF, Greece; ISF, MINERVA, GIF, DIP and Benoziyo Center,
Israel; INFN, Italy; MEXT and JSPS, Japan; CNRST, Morocco; FOM and NWO,
Netherlands; BRF and RCN, Norway; MNiSW, Poland; GRICES and FCT, Portugal; MERYS
(MECTS), Romania; MES of Russia and ROSATOM, Russian Federation; JINR; MSTD,
Serbia; MSSR, Slovakia; ARRS and MIZ\v{S}, Slovenia; DST/NRF, South Africa;
MICINN, Spain; SRC and Wallenberg Foundation, Sweden; SER, SNSF and Cantons of
Bern and Geneva, Switzerland; NSC, Taiwan; TAEK, Turkey; STFC, the Royal
Society and Leverhulme Trust, United Kingdom; DOE and NSF, United States of
America.

The crucial computing support from all WLCG partners is acknowledged
gratefully, in particular from CERN and the ATLAS Tier-1 facilities at
TRIUMF (Canada), NDGF (Denmark, Norway, Sweden), CC-IN2P3 (France),
KIT/GridKA (Germany), INFN-CNAF (Italy), NL-T1 (Netherlands), PIC (Spain),
ASGC (Taiwan), RAL (UK) and BNL (USA) and in the Tier-2 facilities
worldwide.

\begin{flushleft}
{\Large The ATLAS Collaboration}

\bigskip

G.~Aad$^{\rm 48}$,
T.~Abajyan$^{\rm 21}$,
B.~Abbott$^{\rm 112}$,
J.~Abdallah$^{\rm 12}$,
S.~Abdel~Khalek$^{\rm 116}$,
A.A.~Abdelalim$^{\rm 49}$,
O.~Abdinov$^{\rm 11}$,
R.~Aben$^{\rm 106}$,
B.~Abi$^{\rm 113}$,
M.~Abolins$^{\rm 89}$
% ...
\bigskip
\\
% ...
$^{11}$ Institute of Physics, Azerbaijan Academy of Sciences, Baku, Azerbaijan\\
$^{12}$ Institut de F{\'\i}sica d'Altes Energies and Departament de F{\'\i}sica de la Universitat Aut{\`o}noma de Barcelona, Barcelona, Spain\\
% ...
$^{21}$ Physikalisches Institut, University of Bonn, Bonn, Germany\\
% ...
$^{48}$ Fakult{\"a}t f{\"u}r Mathematik und Physik, Albert-Ludwigs-Universit{\"a}t, Freiburg, Germany\\
$^{49}$ Section de Physique, Universit{\'e} de Gen{\`e}ve, Geneva, Switzerland\\
% ...
$^{89}$ Department of Physics and Astronomy, Michigan State University, East Lansing MI, United States of America\\
% ...
$^{106}$ Nikhef National Institute for Subatomic Physics and University of Amsterdam, Amsterdam, Netherlands\\
% ...
$^{112}$ Homer L. Dodge Department of Physics and Astronomy, University of Oklahoma, Norman OK, United States of America\\
$^{113}$ Department of Physics, Oklahoma State University, Stillwater OK, United States of America\\
% ...
$^{116}$ LAL, Universit{\'e} Paris-Sud and CNRS/IN2P3, Orsay, France\\
% ...
$^{d}$ Also at Particle Physics Department, Rutherford Appleton Laboratory, Didcot, United Kingdom\\
% ...
\end{flushleft}

\end{document}